\documentclass[11pt,twoside]{article}
\usepackage{amsmath,amsfonts,amssymb}
\usepackage{latexsym}
\usepackage{dcolumn}
\usepackage{graphicx,epsfig}
\usepackage{amsthm}
\usepackage{color}

\begin{document}

\setcounter{page}{1}

\pagestyle{plain} \vspace{1 cm}

\begin{center}
{\large{\bf { The Modified Quantum Mechanics from the Covariant Generalized Uncertainty Principle}}}\\

\vspace{1 cm}
{\bf Mohaddeseh Seifi$^{\dag,}$}\footnote{m.seifi@stu.umz.ac.ir}\quad and \quad {\bf Akram S. Sefiedgar$^{\dag,}$}\footnote{a.sefiedgar@umz.ac.ir}\\
\vspace{0.5 cm}
$^{\dag}$Department of Theoretical Physics, Faculty of Basic Sciences,\\ University of Mazandaran,\\
P. O. Box 47416-95447, Babolsar, Iran\\

\end{center}

\vspace{1.5cm}

\begin{abstract}
One can use the generalized uncertainty principle (GUP) to incorporate the minimum measurable length in quantum gravity. It may be interesting to have a minimal time interval as well as the minimal length in the relativistic version of quantum mechanics in the presence of the gravitational effects. In this paper, we consider a covariant version of the generalized uncertainty principle to investigate the effects of both the minimal time and the minimal length. Using the covariant GUP, the energy-momentum dispersion relation is modified.  Starting with the modified dispersion relation, the corrections to the wave function are obtained and the problems of the particle in a box and the Hydrogen atom are revisited.
\end{abstract}

\vspace{2cm}
\newpage
\section{Introduction}
The existence of the minimal observable length is a common feature of investigations in string theory and most of the candidates of quantum gravity \cite{Veneziano1986, Amati1987, Amati1989, Gross1987, Konishi1990, Guida1991, Kato1990, Garay1995, Capozziello2000, Maggiore1993, Maggiore1993*, Maggiore1994,Scardigli1999}. Loop Quantum Gravity, Doubly special relativity and black hole physics can predict the emergence of such a minimal observable length. However, the existence of the minimal measurable length is in contradiction with the Heisenberg uncertaity principle (HUP) in quantum mechanics which permits an infinitely small position uncertainty. Hence, it is necessary to add some corrections to the standard uncertainty relation and also to the canonical non-commutative relations \cite{Kempf1995}.  To incorporate the effects of the minimal observable length in quantum gravity, the standard uncertainty relation is replaced by the generalized uncertainty principle \cite{Hossenfelder2013, Adler2010, Tawfik2014,Kempf1994,Kempf1997,Adler1999,Ali2009,Das2009,Ali2011}.

The minimal observable length is normally considered to be at the order of the Planck length, $l_{pl} \approx 10^{-35} m$. It is clear that the minimal length is not a Lorentz invariant quantity. Hence, the theories incorporating the minimal length may break Lorentz covariance. The issue of Lorentz covariance breaking may be a controversial challenge for authors in using GUP models in relativistic theories. However, there are some attempts to propose a Lorentz covariant GUP which reduces to the non relativistic versions at low energies \cite{Quesne2006, Quesne2010, Kober2010, Das2010, Pedram2011, Pramanik2013, Khosropour2016, Deb2016, Todorinov2019,Basilakos2010,Dahab2014,Faizal2016}. 
Of course, it is interesting to investigate the minimal time interval as well as the minimal measurable length. The covariant GUP may provide a suitable framework to investigate the effects of both the minimal time and the minimal length.

In 1947, ُSnyder proposed the concept of the discrete spacetime and abondon the assumption of the continuous spacetime \cite{Snyder1947}. More over, Salecker and Wigner used a quantum clock in measuring the distances. They suggested to register the minimal time \cite{Wigner1957, Salecker1958}. Wilczek and Shapere introduced the term of time crystal when they studied the spontaneous symmetry breaking \cite{Shapere2012, Wilczek2012}. The existence of the minimal time is also investigated in other literatures \cite{Zhang2017, Itzhaki1994, Hooft1996, Bruno2013, Watanabe2015, Castillo2014, Faizal2017}. Therefore, it may be interesting to consider the minimal time as well as the minimal length. For this purpose, the covariant GUP may play an important role. So it seems necessary to consider the time as an observable. Some attempts have been done to view time as a quantum mechanical observable under certaint conditions \cite{Balasubramanian2015, Buscha1994, Olkhovsky2007, Olkhovsky2008, Olkhovsky2009, Brunetti2010} to find the covariant GUP. The covariant GUP can be applied in studying the problems in quantum field theory.

In this letter, the effects of the covariant GUP have been studied in quantum mechanics. A covariant version of GUP has been used to modify the energy-momentum dispersion relation. Then, the modified Klein-Gordon (K-G) equation has been obtained in $1+1$ dimensions. Using the modified Klein-Gordon equation, the problems of the particle in a box and the Hydrogen atom have been investigated.

\section{From the Covariant GUP to the Modified Wave Function}
According to \cite{Faizal2016,Ryder1996, Schwabl2007}, we start with the modified Heisenberg algebra and the modified momentum operator as
\begin{equation}\label{s1}
[x^i,p_j]=i \hbar [\delta^{i}_{j}-\alpha |p^k p_k|^{1/2} \delta^{i}_{j}+\alpha |p^{k}p_{k}|^{-1/2}p^{i}p_{j}+\alpha^2 p^k p_k \delta^{i}_{j}+3\alpha^2 p^i p_j],
\end{equation}
\begin{equation}\label{s2}
p_{i}=-i \hbar (1- \hbar \alpha \sqrt{-\partial^{j} \partial_{j}}-2 \hbar^2 \alpha^2 \partial^{j} \partial_{j})\partial_{i},
\end{equation}
where $i$,$j$,$k$ $\in$ \{1,2,3\}. Note that $\alpha=\alpha_0 / M_{pl}c = \alpha_0 \ell_{pl}/ \hbar$ and $M_{pl}$ is the Planck mass. The parameter $c$ is the light velocity, $\hbar$ is the reduced Planck constant and $\alpha_0 \approx 1$.
Now, we can write the generalized uncertainty principle which is consistent with the modified Heisenberg algebra as  
\begin{align}\label{s3}
\Delta x \Delta p &\geq \frac{\hbar}{2}[1-2\alpha \langle p \rangle +4 \alpha^2 \langle p^2 \rangle] \\  \nonumber
&\geq \frac{\hbar}{2}[1+(\frac{\alpha}{\sqrt{\langle p^2 \rangle}}+4 \alpha^2)\Delta p^2+4 {\alpha^2 \langle p \rangle}^2-2 \alpha \sqrt{\langle p^2 \rangle}].
\end{align}
To incorporate the minimal time as well as the minimal length, we must consider the covariant form of the space-time commutators
\begin{equation}\label{s4}
[x^{\mu},p_{\nu}]=i \hbar [\delta^{\mu}_{\nu}-\alpha |p^{\rho}p_{\rho}|^{1/2} \delta^{\mu}_{\nu}+\alpha |p^{\rho}p_{\rho}|^{-1/2}p^{\mu}p_{\nu}+\alpha^2 p^{\rho} p_{\rho} \delta^{\mu}_{\nu}+3\alpha^2 p^{\mu}p_{\nu}],
\end{equation}
where the modified momentum operator and the position operator are
\begin{equation}\label{s5}
p_{\mu}=-i \hbar (1- \hbar \alpha \sqrt{-\partial^{\nu} \partial_{\nu}}-2 \hbar^2 \alpha^2 \partial^{\nu} \partial_{\nu})\partial_{\mu},
\end{equation}
and
\begin{equation}\label{s6}
x_{\mu}=x_{0{\mu}}.
\end{equation}
In the above, $\mu, \nu \in \{0,1,2,3\}$. Note that $x^{\mu}$ and $p^{\mu}$ are physical position and momentum operators which are not canonical conjugate. The canonical commutation relations are
\begin{equation}\label{s7}
[x_0^{\mu},p_0^{\nu}]=i \hbar \eta^{\mu \nu},
\end{equation}
where $p_0^{\mu}=-i \hbar \frac{\partial}{\partial x_0^{\mu}}$.

It is now possible to modify the energy-momentum dispersion relation. The energy-momentum dispersion relation is written as
\begin{equation}
p^{\mu}p_{\mu}=m^2c^2.
\end{equation}
Substituting equation (\ref{s5}) in the energy-momentum dispersion relation leads to
\begin{equation}\label{3}
p_0^{\mu}p_{0{\mu}}-2 \alpha (p_0^{\mu}p_{0{\mu}})^{3/2}+4 \alpha ^2 (p_0^{\mu}p_{0{\mu}})^2-m^2c^2=0.
\end{equation}
Solving equation (\ref{3}), one can find 
\begin{equation}\label{3+1}
p_0^{\mu}p_{0{\mu}}=m^2c^2+2\alpha ^2 m^4c^4+ \mathcal{O}(\alpha^4),
\end{equation}
which is the modified dispersion relation.

Now, one can write the modified dispersion relation as an operator equation
\begin{equation}\label{3+2}
-\hbar^2 \frac{\partial^2}{\partial t_0^2}\Psi(t_0,\vec{x_0})+\hbar^2 c^2 \vec \nabla_0^2 \Psi(t_0,\vec{x_0})-[m^2c^4+2 \alpha^2m^4c^6]\Psi(t_0,\vec{x_0})=0,
\end{equation}
where $\Psi(t_0,\vec{x_0})$ is a wave function.
It is clear that equation (\ref{3+2}) is the modified  K-G equation which reduces to the standard K-G equation when $\alpha=0$.

In $(1+1$)-dimensional space-time, one can consider $\Psi(t_0,x_0)=T(t_0)U(x_0)$ and write the modified K-G equation as
\begin{equation}\label{3+3}
-\hbar^2 \frac{\partial^2}{\partial t_0^2}T(t_0)U(x_0)+\hbar^2 c^2 \frac{\partial^2}{\partial x_0^2} T(t_0)U(x_0)-[m^2c^4+2 \alpha^2m^4c^6]T(t_0)U(x_0)=0.
\end{equation}
Some manipulations lead to
\begin{equation}\label{3+4}
-\frac{\hbar^2 c^2}{U(x_0)} \frac{\partial^2}{\partial x_0^2}U(x_0)+[m^2c^4+2 \alpha^2m^4c^6]=-\frac{\hbar^2}{T(t_0)}  \frac{\partial^2}{\partial t_0^2} T(t_0).
\end{equation}
It is interesting to attribute the covariant GUP corrections to both the temporal and the spatial sides of the differential equation. Hence, one can write
\begin{equation}\label{3+5}
-\frac{\hbar^2 c^2}{U(x_0)} \frac{\partial^2}{\partial x_0^2}U(x_0)+m^2c^4+2 \lambda \alpha^2m^4c^6=-\frac{\hbar^2}{T(t_0)}  \frac{\partial^2}{\partial t_0^2} T(t_0)-2 \gamma \alpha^2m^4c^6.
\end{equation}
where  $\lambda$ and $\gamma$ are dimensionless parameters and $\lambda+\gamma=1$. Clearly, the two sides of the above equation depend on two different variables. This can only be satisfied if both sides are considered to be constant. Equating both sides of equation (\ref{3+5}) with $E^2$, one can obtain two differential equations. 

The spatial differential equation is written as
\begin{equation}\label{3+6}
-\frac{\hbar^2 c^2}{U(x_0)} \frac{\partial^2}{\partial x_0^2}U(x_0)+m^2c^4+2 \lambda \alpha^2m^4c^6=E^2,
\end{equation}
which can be solved to find the spatial wave function as
\begin{equation}\label{3+7}
U(x_0)=c_1 e^{\frac{i \sqrt{E^2-m^2 c^4-2 \lambda\alpha^2 m^4 c^6}}{ \hbar c}x_0}+c_2 e^{\frac{-i \sqrt{E^2-m^2 c^4-2 \lambda \alpha^2 m^4 c^6}}{ \hbar c}x_0}.
\end{equation}
The temporal differential equation can be written as
\begin{equation}\label{3+8}
-\frac{\hbar^2}{T(t_0)}  \frac{\partial^2}{\partial t_0^2} T(t_0)-2 \gamma \alpha^2m^4c^6=E^2,
\end{equation}
which can be solved to find the temporal wave function as
\begin{equation}\label{7}
T(t_0)=c_3 e^{\frac{i \sqrt{E^2+2 \gamma \alpha^2 m^4 c^6}}{ \hbar }t_0}+c_4 e^{\frac{-i \sqrt{E^2+2 \gamma \alpha^2 m^4 c^6}}{ \hbar }t_0}.
\end{equation}
Clearly, the functions $U(x_0)$, $T(t_0)$ and hence the total wave function have been modified by using the covariant GUP. It means that the probability density and the time evolution of the systems can be corrected.

\section{A Free Particle in a Box}
Considering a free particle in a box with length $L_0$, one can solve  the differential equation (\ref{3+6}) and find a convenient solution as 
\begin{equation}
U(x_0)=A \sin k'x_0+B \cos k'x_0,
\end{equation}
where $k'=\frac{\sqrt{E^2-m^2c^4-2 \lambda \alpha^2 m^4 c^6}}{\hbar c}$. In a box  with length $L_0$, the boundary conditions $U(x_0=0)=0$ and $U(x_0=L_0)=0$ necessitates
\begin{equation}
B=0  \qquad and  \qquad  k'=\frac{n \pi}{L_0} \quad (n=1,2,3,...).
\end{equation}
Hence, one can write the normalized spatial wave function as
\begin{equation}
U(x_0)=\sqrt{\frac{2}{L_0}} \sin \frac{n \pi x_0}{L_0}.
\end{equation}
Now, it is possible to find the energy levels as
\begin{equation}
E= mc^2 (1+\lambda\alpha^2 m^2c^2)+\frac{n^2 \pi^2 \hbar^2}{2mL_0^2}(1- \lambda\alpha^2 m^2c^2)- \frac{n^4 \pi^4 \hbar^4}{8m^3c^2L_0^4}(1-6 \lambda\alpha^2 m^2c^2).
\end{equation}
Note that the total energy includes a GUP corrected rest mass, a GUP corrected non-relativistic energy and a GUP corrected relativistic effects in a box.

It is also interesting to study the temporal analog of the particle in a box. The authors in \cite{Faizal2016} have considered this case and studied the concept of time crystals from the minimum time uncertainty. Here, we are going to use the covariant GUP to study a free particle in a temporal box.
Starting with the differential equation (\ref{3+8}), one can find a convenient solution as
\begin{equation}
T(t_0)=C \sin k''t_0+D \cos k''t_0,
\end{equation}
where $k''=\frac{\sqrt{E^2+2\alpha^2 \gamma m^4 c^6}}{\hbar}$.
In a temporal box  with time interval $T_0$, the boundary conditions necessitates
\begin{equation}
D=0 \quad  and \quad k''=\frac{q \pi}{T_0} \quad (q=1,2,3,...).
\end{equation}
As a result, the temporal part of the wave function can be written as
\begin{equation}
T(t_0)=\sqrt{\frac{2}{T_0}} \sin \frac{q \pi t_0}{T_0}.
\end{equation}
Obviously, one can find
\begin{equation}
T_0= \frac{q \pi \hbar}{\sqrt{{E}^2+2 \alpha^2 \gamma m^4c^6}},
\end{equation}
which refers to the existence of the discrete time instead of the continuous one. 

The particle in a box can play the role of a measurement instrument which can provide a spatial scale at the order of $L_0$ and a temporal scale at the order of $T_0$. To probe the Planckian space-time cells, it is necessary to decrease $L_0$ and $T_0$ which can be reached only by increasing energy. In other words, one can demonstrate the space-time discreteness only in high energy regime. 

\section{Relativistic Hydrogen Atom}
The wave function $\Psi(t_0,\vec{r_0})$ for an electron at a radial distance $r_0$ from the proton in a Hydrogen atom can be derived from the modified K-G equation \cite{Naudts2005, Ducharme2010}
\begin{equation}\label{8+1}
-\frac{\partial^2}{\partial t_0^2}\Psi(t_0,\vec{r_0})+ c^2 \vec \nabla_0^2 \Psi(t_0,\vec{r_0})-\frac{c^2}{\hbar^2}(m^2c^4+2 \alpha^2m^4c^6)\Psi(t_0,\vec{r_0})=0.
\end{equation}
Considering a minimal coupling to a magnetic field, one can write
\begin{equation}\label{8+2}
i\hbar \frac{\partial}{\partial x_0^\mu} \longrightarrow i\hbar \frac{\partial}{\partial x_0^\mu}- \frac{e}{c}A_\mu,
\end{equation}
where $A^\mu =(\phi,\vec{A})$ is the four-potential. Note that $\phi$ and $\vec{A}$ are scalar and vector potentials respectively. The potential is considered as the Coulomb potential 
\begin{equation}\label{8+3}
V=-\kappa\frac{\hbar c}{r_0},
\end{equation}
where $\kappa= \frac{e^2}{4\pi\epsilon_0 \hbar c}$ is the fine structure constant.
Substituting equations (\ref{8+2}) and (\ref{8+3}) into (\ref{8+1}), one can write
\begin{equation}\label{8+4}
-(i\frac{\partial}{\partial t_0}+\kappa\frac{ c}{r_0})^2 \Psi-c^2 \nabla_0^2 \Psi+\frac{c^2}{\hbar^2}(m^2c^4+2\alpha^2m^4c^6)\Psi=0.
\end{equation}
The wave function in the spherical polar coordinates can be written as
\begin{equation}\label{8+5}
{\Psi}_{nlm}={R}_{nl}(r_0){Y}_{lm}(\theta,\phi) e^{-i\frac{ \sqrt{E^2+2 \gamma \alpha^2 m^4 c^6}}{ \hbar }t_0}.
\end{equation}
The eigenvalue equation for the orbital angular momentum operator is
\begin{equation}\label{8+7}
L^2{Y}_{lm}=l(l+1){Y}_{lm}.
\end{equation}
Using equations (\ref{8+5}) and (\ref{8+7}), one can write equation (\ref{8+4}) as
\begin{equation}\label{8+8}
\left[\frac{1}{r_0^2} \frac{\partial}{\partial r_0}(r_0^2\frac{\partial}{\partial r_0})+\frac{D^2}{c^2\hbar^2}+\frac{\kappa^2}{r_0^2}+\frac{2D}{c\hbar}\frac{\kappa}{r_0}-\frac{l(l+1)}{r_0^2}-\frac{1}{\hbar^2}(m^2c^2+2\alpha^2m^4c^4)\right]{R}_{nl}=0,
\end{equation}
in which the definition $D^2=E^2+2\gamma \alpha^2m^4c^6$ is used for simplicity.
The solution of equation (\ref{8+8}) is written as
\begin{equation}\label{8+9}
{R}_{nl}(r_0)=\frac{{N}_{nl}}{{r_0}^{\eta_l}}\exp(-\frac{r_0}{{r_0}_{nl}})\sum a_s r_0^s,
\end{equation}
where
\begin{equation}\label{8+10}
\eta_l=\frac{1}{2}\pm \sqrt{(l+\frac{1}{2})^2-\kappa^2},
\end{equation}
and the parameters $N_{nl}$, $r_{nl}$ and $a_s$ are constants. Substituting equation (\ref{8+9}) and (\ref{8+10}) into equation (\ref{8+8}) and some manipulations lead to
$$\sum a_s [\frac{s(s-1){r_0}^{s-2}}{{r}_{0nl}^2} +((1-\eta_l)\frac{{r_0}_{nl}}{r_0}-1)\frac{2s{r_0}^{s-1}}{{r_0}_{nl}^2}]+$$
\begin{equation}\label{8+11}
[\frac{D^2}{c^2\hbar^2}+\frac{2D}{c\hbar}\frac{\kappa}{r_0}-\frac{2(1-\eta_l)}{r_0{r_0}_{nl}}+\frac{1}{{r_0}_{nl}^2}
-\frac{1}{\hbar^2}(m^2c^2+2\alpha^2m^4c^4)]\sum a_sr_0^s=0.
\end{equation}
Solving the above equation yields
\begin{equation}\label{8+12}
{r}_{0nl}=\frac{\hbar c}{\sqrt{m^2c^4+2\alpha^2 m^4c^6-D^2}},
\end{equation}
and
\begin{equation}\label{8+13}
D=\frac{\hbar c(n+1-\eta_l)}{\kappa {r}_{0nl}}.
\end{equation}
From equations (\ref{8+12}) and (\ref{8+13}), the energy eigenvalues for Hydrogen atom can be obtained
\begin{eqnarray}\label{8+14}
E=mc^2 [1+\alpha^2m^2c^2+\gamma\alpha^2m^2c^2(1-\frac{\kappa^2}{(n+1-{\eta}_{nl})^2})- \frac{\kappa^2}{2(n+1-{\eta}_{nl})^2}\nonumber\\ -\frac{\alpha^2m^2c^2\kappa^2}{2(n+1-{\eta}_{nl})}- \frac{\gamma\alpha^2m^2c^2\kappa^2}{2(n+1-{\eta}_{nl})}(1- \frac{\kappa^2}{(n+1-{\eta}_{nl})^2})].
\end{eqnarray}
The covariant GUP corrected energy eigenvalues of a relativistic Hydrogen atom has been obtained. It is interesting to attribute the GUP corrections into both the spatial and temporal wave functions. In the case $\alpha=0$, the result reduces to the energy levels of the relativistic Hydrogen atom in \cite{Naudts2005, Ducharme2010}. 

\section{Conclusions}

The existence of the minimal time interval has been studied in some literatures. The possibility of the existence of the minimal time has motivated us to consider the generalized uncertainty principle in a covariant form. The covariant GUP may provide a framework to investigate the effects of both the minimal time and the minimal measurable length. Using the covariant GUP, we have modified the energy-momentum dispersion relation in a relativistic model. Then, we have derived the modified K-G equation. Attributing the GUP corrections to the temporal wave function as well as the spatial one, we have solved the modified K-G equation and derived the corrections to the total wave function. As a result, the time evolution and the probability density can be corrected. Finally, two important problems of the free particle in a box and the Hydrogen atom have been studied.
We have considered the free particle in a $(1+1)$-dimensional box with spatial length $L_0$ and the time interval $T_0$. The modified wave function of the particle and its energy spectrum have been obtained. It is important to point that the rest mass has been corrected by the effects of GUP which can be detected in the future high energy experiments. We have also found that time can be a discrete quantity and its discreteness is important in high energies.   
Then, the wave function of the relativistic Hydrogen atom has been obtained with GUP corrections. The energy levels of the Hydrogen atom have been derived. The GUP corrections to the rest mass have been presented. The results can be reduced to the ones in other works in the absence of GUP effects. It is expected that the covariant GUP corrections to the temporal wave function lead to important effects in the quantum systems with the time dependent potential.

\end{document}